# Relativistic Electron Acceleration and the 'Ankle' Spectral Feature in Earth's Magnetotail Reconnection

Weijie Sun[1], Mitsuo Oka[1], Marit Øieroset[1], Drew L. Turner[2], Tai Phan[1], Ian J. Cohen[2], Xiaocan Li[3], Jia Huang[1], Andy Smith[4], James A. Slavin[5], Gangkai Poh[6,7], Kevin J. Genestreti[8,9], Dan Gershman[6], Kyunghwan.Dokgo[8], Guan Le[6], Rumi Nakamura[10], James L. Burch[8]

Corresponding author: Weijie Sun, weijiesun@berkeley.edu

[1], Space Sciences Laboratory, University of California, Berkeley, CA, USA

[2], Space Exploration Sector, Johns Hopkins Applied Physics Laboratory, Laurel, MD, USA.

[3], Department of Physics and Astronomy, Dartmouth College, Hanover, NH, USA

[4], Department of Mathematics, Physics and Electrical Engineering, Northumbria University, Newcastle Upon Tyne, UK

[5], Department of Climate and Space Sciences and Engineering, University of Michigan, Ann Arbor, MI, USA.

[6], NASA Goddard Space Flight Center, Greenbelt, MD, USA

[7], Center for Research and Exploration in Space Sciences and Technology II, Catholic University of America, Washington, DC, USA

[8], Southwest Research Institute, San Antonio, TX, USA

[9], University of New Hampshire, Durham, NH, USA

[10], Space Research Institute, Austrian Academy of Sciences, Graz, Austria

**Abstract** (<250 words)

Electrons are accelerated to high, non-thermal energies during explosive energy-release events in space, such as magnetic reconnection. However, the properties and acceleration mechanisms of relativistic electrons directly associated with reconnection X-line are not well understood. This study utilizes Magnetospheric Multiscale (MMS) measurements to analyze the flux and spectral features of sub-relativistic to relativistic (~ 80 to 560 keV) electrons during a magnetic reconnection event in Earth's magnetotail. This event provided a unique opportunity to measure the electrons directly energized by X-line as MMS stayed in the separatrix layer, where the magnetic field directly connects to the X-line, for approximately half of the observation period. Our analysis revealed that the fluxes of relativistic electrons were clearly enhanced within the separatrix layer, and the highest flux was directed away from the X-line, which suggested that these electrons originated directly from the X-line. Spectral analysis showed that these relativistic electrons deviated from the main plasma sheet population and exhibited an "ankle" feature similar to that observed in galactic cosmic rays. The contribution of "ankle" electrons to the total electron energy density increased from 0.1% to 1% in the separatrix layer, though the spectral slopes did not exhibit clear variations. Further analysis indicated that while these relativistic electrons originated from the X-line, they experienced a non-negligible degree of scattering during transport. These findings provide clear evidence that magnetic reconnection in Earth's magnetotail can efficiently energize relativistic electrons directly at the X-line, providing new insights into the complex processes governing electron dynamics during magnetic reconnection.

# 1. Introduction

Magnetic reconnection is a fundamental plasma process, where magnetic energy is efficiently dissipated and converted into various forms of particle energy, including kinetic, thermal and non-thermal energies (e.g., Gonzalez & Parker 2016; Ji et al. 2022; Yamada, Kulsrud, & Ji 2010). Its significance extends across diverse space and astrophysical plasma environments, such as magnetic storms (e.g., Angelopoulos et al. 2020; Beyene & Angelopoulos 2024; Gonzalez et al. 1989), substorms (e.g., Angelopoulos et al. 2008; Baker et al. 1996), solar flares (e.g., Demoulin et al. 1993; Shibata 1998) and phenomena associated with magnetars (e.g., Duncan & Thompson 1992; Lyutikov 2003) and pulsars (e.g., Contopoulos 2005; Hakobyan, Philippov, & Spitkovsky 2023; Lyubarsky & Kirk 2001), see a review of Gershman et al. (2024).

Magnetic dissipation actively occurs in the electron and ion diffusion regions of magnetic reconnection, where particles become unmagnetized (e.g., Burch et al. 2016b; Gonzalez & Parker 2016; Sonnerup 1979). These diffusion regions contain non-ideal electric fields that energize particles, as directly measured by NASA's Magnetospheric Multiscale (MMS) mission (e.g., Burch, et al. 2016b; Phan et al. 2018; Torbert et al. 2018). Downstream of the diffusion regions, additional processes could further energize particles. These processes include magnetic curvature-driven Fermi acceleration (e.g., Dahlin, Drake, & Swisdak 2014; Drake et al. 2006; Li et al. 2017), parallel electric potential trapping near the separatrices (e.g., Egedal, Daughton, & Le 2012), and local betatron acceleration (e.g., Hoshino et al. 2001; Jiang et al. 2021; Ma et al. 2024; Sun et al. 2022; Zhong et al. 2020).

Electrons with energies of ~ 300 keV or higher have been observed in the reconnecting current sheet, i.e., near the X-line, in Earth's magnetotail (e.g., Genestreti et al. 2023; Øieroset et al. 2002; Qi et al., 2024) and in solar flares (Chen et al. 2020). These electrons with energies comparable to the electron rest energy ($m_e c^2$ ~ 511 keV, $m_e$ representing the electron rest mass and $c$ being the speed of light), are often termed as relativistic electrons. The spectral features and flux variations of these relativistic electrons, especially their direct connection to the X-line, are still unclear and have not been fully observed.

In Earth's plasma sheet, the main electron spectrum contains a thermal component (with energies below a few keV) and a supra-thermal component (with energies above a few keV), which is often modelled using a Kappa distribution (e.g., Christon et al. 1991; Olbert 1968; Vasyliunas 1968). Recently, Oka et al. (2022) have reported an excess flux of sub-relativistic electrons at the higher energy end (>100 keV) of the spectrum during reconnection events. These sub-relativistic electrons clearly deviate from the plasma sheet main component. However, their analysis focuses on the main component, and the detailed properties of the excess component remain unclear.

In this study, we utilize MMS observations to analyze a magnetic reconnection event in the near-Earth tail. During this event, electrons with energies from sub-relativistic (~ 100 keV) to relativistic (~ 560 keV) appear and deviate from the main plasma sheet spectra. We investigate the flux variations, spectral features and energy density of these electrons. During this event,

MMS remain in the separatrix layer for approximately half of the observation period. Since the magnetic fields in the separatrix layer directly connect to the X-line, we are able to investigate relativistic electrons originating directly from the X-line.

## 2. Magnetic Reconnection in Near-Earth Tail on 2 June, 2018

### 2.1. Data Sources and Instrumentations

This study utilizes field and particle measurements from the MMS mission (Burch et al. 2016a). These measurements include magnetic field data from FluxGate Magnetometers (FGM) (Russell et al. 2016), electron and ion data from the Fast Plasma Investigation (FPI, ~ 0.01 to 30 keV/q) (Pollock et al. 2016), electron data from Fly's Eye Electron Proton Spectrometer (FEEPS, ~ 40 to 600 keV/q) (Blake et al. 2016), and ion data from the Hot Plasma Composition Analyzers (HPCA, <40 keV/q) (Young et al. 2016).

During the reconnection event, the burst-mode measurements from the instruments are available: FGM provides magnetic field vectors at ~ 128 vectors/s; FPI provides 3D distributions of electrons and ions at 30 millisecond (ms) and 150 ms, respectively; FEEPS provides energetic electron distributions at ~ 0.31 s; and HPCA provides measurements at ~ 10 s.

### 2.2. Overview of Magnetic Reconnection Event from MMS measurements

On 2 June 2018, between ~ 18:27 and 18:34 Coordinated Universal Time (UTC), MMS was located in Earth's plasma sheet at ~ (-13.8, -7.6, 3.1) Earth radii ($R_E$) in Geocentric Solar Magnetospheric (GSM) coordinates. Figure 1 shows observations from MMS1, in which magnetic field vectors and plasma velocities are transformed into the local coordinate of the neutral sheet. In this local coordinate, $\hat{L}$, which is analogous to $\hat{x}$ in GSM, is [0.873, 0.481, -0.08], $\hat{M}$ is [-0.397, 0.797, 0.456] and $\hat{N}$, the normal direction of the neutral sheet, is [0.283, -0.366, 0.887] (see Appendix A for details on how the local coordinate was established). The normal component of the magnetic field ($B_N$) reversed from negative to positive (Figure 1a), coinciding with a reversal of the ion bulk velocity ($V_L$) from tailward to Earthward (Figure 1e). This indicates that MMS traversed a reconnection X-line from tailward to Earthward, as schematically illustrated in Figure 1i.

At ~ 18:27:00 UTC, MMS was located southward near the neutral sheet with $B_L$ ~ - 10 nT and the plasma $\beta$ (the ratio of plasma thermal pressure to magnetic pressure, not shown) exceeding 10. MMS began observing a tailward ion flow with a speed of ~ 200 km/s and a strong velocity component along $\hat{M}$ (i.e., mostly duskward, Figure 1e). Simultaneously, the derived lobe magnetic field intensity (Figure 1h), based on pressure balance between the plasma sheet and the lobe, started to decrease. From ~ 18:27:00 to 18:28:30 UTC, electrons with energies around 1 keV exhibited higher fluxes moving towards the X-line (anti-parallel to the magnetic field, Figure 1g) than away from it (parallel to the magnetic field). While electrons with energies above ~ 2 keV had higher fluxes moving away from the X-line than towards it (Figure 1f). These counter-streaming electrons suggested that MMS was located in the separatrix layer (e.g., Nagai et al. 2001).

After exiting the separatrix layer, MMS moved closer to the neutral sheet ($|B_L| < 5$ nT) but the ion bulk velocity remained low until ~ 18:29:10 UTC. Then MMS moved southward, as revealed by the gradual increase in $|B_L|$. Meanwhile, the ion bulk velocity began to increase significantly, reaching 550 km/s at ~ 18:29:30 UTC. Between ~ 18:29:55 and 18:31:10 UTC, MMS re-entered the separatrix layer, as evidenced by the reappearance of counter-streaming electrons.

At ~ 18:30:49 UTC, MMS entered the Earthward side of the X-line. It then traversed the neutral sheet three times before entering the separatrix layer on the northern side of the plasma sheet at ~ 18:32:15 UTC. During this reconnection event, MMS spent around half of the observation time within the separatrix layer. At least one ion-scale flux rope and four travelling compression regions (TCRs) (e.g., Slavin et al. 1992), a signature of travelling flux ropes in the plasma sheet center, were observed during this event.

### 2.3. Relativistic Electron Variations During Magnetic Reconnection Event

Electron particle fluxes exhibited distinct energy-dependent variations during this reconnection event, as shown in Figure 2. During the low bulk velocity interval from ~ 18:27:00 to 18:29:30 UTC (hereafter refer to as pre-active phase), electron fluxes for energies below 100 keV slightly decreased, while high-energy electrons (~220 keV) increased. At ~ 18:29:30 UTC when the ion bulk velocity significantly increased, which referred to as the active phase, a sharp enhancement of electron fluxes for energies above ~ 50 keV appeared. This enhancement exhibited an energy dispersion with higher energy electrons appearing earlier, which is analyzed in sections 2.5.

From ~ 18:29:30 to 18:32:00 UTC, MMS resided near or within the separatrix layer and traversed the primary X-line from tailward to Earthward. $|B_L|$ increased gradually to ~ 30 nT (Figure 2a) and low energy electron fluxes (<15 keV) decreased together with plasma density. However, electron fluxes with energies up to 560 keV increased (Figures 2e to 2i). As shown in the pitch angle distributions (Figures 2f to 2i), relativistic electron fluxes were enhanced in all directions relative to the magnetic field. However, the electrons moving parallel to the magnetic field contained the highest fluxes when MMS was tailward of the X-line. After MMS traversed the X-line (~18:30:49 UTC) from tailward into Earthward, electron fluxes moving anti-parallel became the highest. Since MMS was southward of the neutral sheet, these features of the relativistic electrons aligned with the scenario that they originated from the X-line.

The highest fluxes corresponded to electrons moving away from the X-line, with a slight asymmetry favoring the tailward side of the X-line. On the Earthward side, electrons originating from the X-line tended to be more easily reflected back due to Earth's dipole magnetic field than those on the tailward side, which could explain the weaker asymmetry on the Earthward side.

### 2.4. Density and Energy Partitions of Thermal, Supra-thermal and "Ankle" Components

We utilized the Kappa distribution function (e.g., Olbert 1968; Vasyliunas 1968) to characterize the electron spectra. Figures 3a to 3c show the modeling for an electron spectrum between 18:29:36.6 to 18:29:39.67 UTC. The electron spectrum was modeled using two Kappa distribution functions: one for main plasma sheet electrons with energies ranging from ~ 0.1 to 70 keV (e.g., Christon, et al. 1991) and the other for sub-relativistic (~ 100 keV) to relativistic (~ 560 keV) electrons. These sub-relativistic and relativistic electrons clearly deviated from the main component, as indicated by a change in the spectral slope from steeper to flatter. They are termed as the "ankle" component because its spectral feature resembles the "ankle" in the galactic cosmic ray energy spectrum. However, the galactic cosmic ray "ankle" is observed at ~ $10^{18}$ eV (e.g., Tanabashi et al. 2018), it is only at ~ $10^5$ eV in our event.

The Kappa distribution function for the electron phase space density ($f_e^\kappa$) is expressed as:

$$f_e^\kappa = \frac{n_{\kappa e}}{2\pi(\kappa \omega_{\kappa e}^2)^{3/2}} \frac{\Gamma(\kappa+1)}{\Gamma(\kappa-1/2)\Gamma(3/2)} \left(1 + \frac{v^2}{\kappa \omega_{\kappa e}^2}\right)^{-\kappa-1} \quad (1)$$

, where $n_{\kappa e}$ represents the number density, $\kappa$ denotes the Kappa value, $\Gamma$ stands for the Gamma function, $\omega_{\kappa e}$ represents the thermal velocity, and $v$ is the kinetic velocity. The generalized temperature ($T_{\kappa e}$) is expressed as

$$\kappa_B T_{\kappa e} = \frac{\omega_{\kappa e}^2 \kappa m_e}{2\kappa - 3} \quad (2)$$

, where $\kappa_B$ represents the Boltzmann constant.

Following the definition by Oka et al. (2015), we use the most probable velocity to further divide the main component into the thermal and supra-thermal components. The thermal temperature is defined by an adjusted Maxwellian distribution:

$$T_M = \left(\frac{\kappa - 3/2}{\kappa}\right) T_{\kappa e} \quad (3)$$

, and the thermal density is:

$$n_M = 2.718 \frac{\Gamma(\kappa+1)}{\Gamma(\kappa-1/2)} \kappa^{-3/2} \left(1 + \frac{1}{\kappa}\right)^{-\kappa-1} n_{e\kappa} \quad (4)$$

The supra-thermal component is the difference between the Kappa and adjusted Maxwellian distributions.

In the case of Figures 3a to 3c, thermal component electrons contained a density of 0.41 cm$^{-3}$ and an energy density ($n_M \kappa_B T_M$) of 0.055 nPa. Supra-thermal component had a density of 0.06 cm$^{-3}$ and an energy density of 0.027 nPa. "Ankle" component contained a density of $4.04 \times 10^{-6}$ cm$^{-3}$ and an energy density of $3.36 \times 10^{-5}$ nPa. Thus, the thermal component accounted for ~ 87% of the density and 67% of the energy density. The supra-thermal component contributed ~ 13% of the density and 33% of the energy density. The "ankle" component contributed ~ $8.6 \times 10^{-4}$% of the density and 0.04% of the energy density.

Figures 3d and 3e present three electron spectra: one from the pre-active phase and two from active phase. Of the two from the active phase, one is during the energy dispersion interval, and the other is in the separatrix layer. Through comparing the three spectra, the phase space densities of relativistic electrons were clearly enhanced during the active phase and especially in the separatrix layer. Additionally, the $\kappa$ i.e., the spectral slope, for the supra-thermal and "ankle" components in the separatrix layer were the lowest.

Figure 4 summarizes the modelling results. During the pre-active phase, the number and energy densities of thermal, supra-thermal and "ankle" components decreased as MMS moved from the plasma sheet center to the outer plasma sheet. However, the density and energy density fractions for each component remained relatively stable. Thermal components contributed ~ 89% of the density and ~ 72% of the energy density, supra-thermal components contributed ~ 11% and ~ 28%, and "ankle" components contributed ~ $6.0 \times 10^{-3}$% and ~0.06%, respectively. The spectral slopes for the supra-thermal and "ankle" components decreased slightly. Note that the spectral slopes for the "ankle" components were also obtained using a power law.

After ~ 18:29:30 UTC, during the active phase, the density fractions of thermal components decreased from ~89% to below 81%, and their energy density fractions decreased from ~70% to 51%. The density fractions of supra-thermal components increased from ~11% to 19%, and their energy density fractions increased from 30% to 49%. The maximum values of their energy density fraction coincided with the separatrix layer between 18:30 to 18:31 UTC. Although the contributions from "ankle" components were small, they exhibited large variations. The intensity and fraction of the energy density of the "ankle" components could increase by an order of magnitude. For example, from ~ 18:29:50 to 18:30:10 UTC, coinciding with the separatrix layer, the energy density fraction of "ankle" components increased from ~ 0.08% to 0.9%. The spectral slopes for the supra-thermal components were harder (or flatter) in the separatrix layer than in other regions, while the spectral indices for the "ankle" components were comparable both inside and outside the separatrix layer.

The energy range of the measurements for the "ankle" component was narrow, which normally spanned from ~ 80 to 560 keV in our case. Therefore, when the energy range for the "Ankle" component, after removing the plasma sheet main component, was too narrow (e.g., ~ 200 to 560 keV in our case), we stopped modelling the "ankle" component using Kappa distribution function. Instead, we applied a power law to model the spectra of the "ankle" component without requiring a specific energy range. We note that other functions, such as a power law with an exponential rollover (Ellison & Ramaty 1985) or a Gaussian (Oka et al., 2022), could also model the "ankle" component fairly well. Due to the limited energy range in measurements, it is difficult to definitely determine which distribution function best describes the "ankle" component. This issue is a common concern in the study of energy spectra, as discussed in Oka et al. (2018).

2.5. Energy Dispersion of Relativistic Electrons Associated with Ion Bulk Velocity Enhancement

Figure 5 shows observations of the dispersed electron flux enhancement starting at ~ 18:29:30 UTC, which was accompanied by the enhancement of ion bulk velocity. Electrons with higher energy appeared earlier than lower energy electrons. In Figure 5c, the asterisks indicate the times when electron fluxes begin to clearly increase in each energy channel. This energy dispersion is often interpreted as a time-of-flight effect (e.g., Bai et al. 2019), and fitting the starting times of the flux increases can obtain the source distance.

During the dispersion, the spacecraft first moved towards the neutral sheet and then moved away from it, which was revealed by $|B_L|$ decreasing first and then increasing. Only the dispersion that occurred while MMS moved towards the neutral sheet was considered in the following analysis. The energy dispersion was assumed to be a spatial feature. If we further assumed electrons predominately travelled along field lines, then the distance between the MMS and source location could be determined from the following equation:

$$\frac{1}{V_{para}(t)} = \left(\frac{V_{sc}}{V_c}\right)\frac{1}{L}(t - t_0) \tag{5}$$

, where $V_{sc}$ represents the velocity of spacecraft relative to the plasma sheet, $V_c$ is the convection velocity of field lines in the plasma sheet reference frame, $L$ stands for the distance between the MMS and source locations, $t$ represents the time when MMS observe the enhancement of specific energy electrons, and $t_0$ denotes the time when the electrons left the source location. $V_{para}(t)$ represents the parallel velocity of the electrons, which is expressed as follows:

$$V_{para}(t) = c\sqrt{1 - \left(\frac{m_e c^2}{E_k(t) + m_e c^2}\right)^2} \tag{6}$$

, where $E_k$ represents the kinetic energy of electrons.

In this case, $V_{sc}$ was ~ -40 km/s, which was obtained by averaging the velocity from 18:29:33 to 18:29:40 UTC in the $\widehat{N}$ direction applying the spatio-temporal difference method (Shi et al. 2006). $V_c$ was ~ -205 km/s, which was obtained by projecting the ion bulk velocity into the $\widehat{N}$ and subtracting the velocity of plasma sheet motion. The slope obtained from the fitting $t$ and $1/V_{para}(t)$ in Figure 5d was ~ $2.3 \times 10^{-3} \pm 7.8 \times 10^{-4}$ $R_E^{-1}$. Thus, the distance $L$ between MMS and the source location was estimated to be ~ $71 \pm 24.0$ $R_E$. This distance was larger than the half width of the magnetotail (~ 20 $R_E$) at $X_{GSM}$ ~ -14 $R_E$ (e.g., Shue et al. 1998). We compared the distance $L$ with the cross-tail width of the magnetotail as because the main component of the magnetic field during this period was $B_M$.

One of the assumptions in the equation (5) was that electrons experienced negligible pitch angle scattering from the source location to MMS. However, within the plasma sheet close to the X-line, magnetic curvature-related pitch angle scattering could be strong due to the larger gyro-radii of relativistic electrons and smaller magnetic curvature radii. Furthermore, electron transport involved magnetic and electric field fluctuations, i.e., non-adiabatic processes, within the plasma sheet (e.g., Imada, Hoshino, & Mukai 2008; Wang et al. 2006). During the energy dispersion, flux enhancements occurred in all directions relative to the magnetic field, although electrons moving away from the X-line exhibited the highest fluxes. This feature was consistent with the idea that non-adiabatic processes were important during the transport of these electrons.

### 3. Discussion on Origins and Energization of the "Ankle" Component

The "ankle" electrons, which deviated from the main plasma sheet component in electron spectra, was modelled using a separate Kappa distribution function. The number and energy densities of the "ankle" electrons remained relatively stable during the pre-active interval when ion bulk velocity was small. In contrast, the number density and energy density could vary by an order of magnitude during the active interval when ion bulk velocity was large.

The fluxes of relativistic electrons showed a dropout between ~ 18:30:40 to 18:30:50 UTC, which might indicate that MMS moved out of the reconnection outflow region and into the inflow region. In that region, $B_L$ was ~ 28 nT, and the electron number density was ~ 0.32 cm$^{-3}$, which corresponded to an Alfvén speed of ~ 1100 km/s. Since the ion bulk velocity in our event did not exceed 1000 km/s, it was likely that magnetic reconnection only dissipated magnetic field and energized electrons inside the plasma sheet. This indicated that an "ankle" component might already exist in the plasma sheet before magnetic reconnection occurred. It was because of their low flux, lower than the one-count-level of FEEPS, they could not be distinguished during the quite periods. However, magnetic reconnection energized these electrons and enhanced their fluxes in the electron spectra above the one-count-level of FEEPS. Nevertheless, we still could not exclude the possibility that they were directly formed due to the energization from magnetic reconnection.

The relativistic electrons might not be locally energized. Their enhancements parallel to the local magnetic field and overlapping with the separatrix suggested that the energization processes such as parallel potential along the separatrices (e.g., Egedal, et al. 2012) could still play a role. However, magnetic curvature-driven Fermi-type and betatron accelerations (e.g., Drake et al., 2006; Dahlin et al., 2014; Li et al. 2019), which were typically important downstream of the reconnection diffusion regoin, were unlikely to be significant for energizing the relativistic electrons in the separatrix layer.

## 4. Summary

We have investigated the fluxes, spectral properties and energy densities of the sub-relativistic to relativistic electrons (~ 80 to 560 keV) during a magnetic reconnection event in the near-Earth tail. In this event, MMS stayed in the separatrix layer for approximately half of the observation time. The relativistic electron fluxes were strongest in the separatrix layer, and there were more electrons moving out from the X-line, indicating that they originate from the X-line.

Relativistic electrons deviate from the main plasma sheet component and can be fitted by a separate Kappa distribution function, which we term as "ankle" component. The "ankle" component contributes ~ $10^{-3}$% of the number density and 0.01% to 1% of the electron energy density. In the separatrix layer during the high ion bulk velocity interval, the intensity and fraction of the energy density of the "ankle" component increase by approximately an order of magnitude. Meanwhile, the fractions of density and energy density of supra-thermal electron components also significantly increase. The spectral slopes for the supra-thermal components are harder in the separatrix layer than in other regions, while the spectral indices for the "ankle" components are comparable both inside and outside the separatrix layer. The high ion bulk velocity is associated with a stronger reconnection electric field, which appears to more efficiently energize electrons into supra-thermal and relativistic energy ranges.

The increase in ion bulk velocity is accompanied by a sharp enhancement of relativistic electrons. The enhancement shows energy dispersion, i.e., higher energy electrons appearing earlier than lower energy electrons. We suggest that the energy dispersion and enhancements are influenced by time-of-flight effects and pitch angle scattering processes. This is evidenced by the fact that although electrons originating from the X-line exhibit higher fluxes than in other directions, electron fluxes increase in all directions relative to the magnetic field.

## Appendix A: Establishment of the Local Coordinate System of the Neutral Sheet

During the reconnection event from approximately 18:27 to 18:34 UTC, MMS did not traverse the center of the reconnecting neutral sheet. Thus, to establish the local coordinate system of the neutral sheet, we utilized the adjacent neutral sheet crossing of MMS between 18:25 and 18:27 UTC.

We first applied MVA, known as the minimum (or maximum) variance analysis (Sonnerup & Cahill Jr. 1967; Sonnerup & Scheible 1998), to the magnetic field measurements measured by MMS1 between 18:25 to 18:27 UTC. MVA yielded three orthogonal eigenvectors, with each corresponding to an eigenvalue. The ratio between the maximum and intermediate eigenvalues was approximately 195, suggesting a good separation between $\hat{n}_{max}$ and $\hat{n}_{int}$. However, the ratio between the intermediate and minimum eigenvalues was approximately 1.5, indicating less distinct separation between $\hat{n}_{int}$ and $\hat{n}_{min}$. Therefore, the eigenvector $\hat{n}_{max}$, [-0.875, -0.479, 0.075], aligning with the direction of the largest magnetic field direction was well determined.

To determine the normal direction of the neutral sheet, we employed the minimum directional derivative or difference (MDD) method (Shi et al. 2005, 2019), which required simultaneous magnetic field measurements from the four spacecraft of MMS. The MDD technique solved the matrix product of $(\nabla \vec{B})$ and its transpose $(\nabla \vec{B})^T$, where $(\nabla \vec{B})$ represents the magnetic gradient tensor. The MDD technique provided eigenvalues and orthogonal eigenvectors in a time series. The normal direction determined by MDD was averaged over the interval from 18:25:00 to 18:26:30 UTC and denoted as $\hat{N}$, [0.283, -0.366, 0.887]. The $\hat{N}$ was 90.3° from the $\hat{n}_{max}$ determined by MVA. We then derived the $\hat{M}$, [-0.397, 0.797, 0.456], direction according to the right-handed coordinate system. At last, we derived the $\hat{L}$, which was [0.873, 0.481, -0.08].

This hybrid approach for determining the local coordinate of the magnetic structures has been employed in various MMS studies related to the reconnecting neutral sheet, such as Denton et al. (2018); Genestreti et al. (2018); Sun, et al. (2022) etc.

## Appendix B: Modelling of Electron Spectra

Prior to modelling the electron spectra in MMS's measurements, we applied a base-10 logarithm to both the Kappa distribution function and the electron phase space densities. We utilized the Levenberg-Marquardt method (Moré 1977) for nonlinear least-squares curve fitting, implemented via *Mathworks* (2022).

In the modelling, we incorporated the data points from both FPI and FEEPS with energies exceeding 0.1 keV to exclude possible influence from photoelectrons (Gershman et al. 2017). Firstly, we required that the measurements from the highest energy channel from FEEPS contained particle fluxes higher than the background values. Secondly, we averaged the electron measurements over a time window of ~ 3 seconds, including 10 measurements from FEEPS and approximately 100 measurements from FPI. These average values need to exceed the values derived from one-count for FPI and the background for FEEPS. The first time interval satisfying the above requirements and the start time of Figure 4 is ~ 18:28:02.82 UTC.

Since we employed two Kappa distribution functions to model the electron spectra, the reduced $\chi^2$ was calculated as follows:

$$\chi^2 = \sum_{i=1}^{N} \left(\log_{10} f_e(i) - \log_{10}\bigl(f_e^{\kappa 1}(i) + f_e^{\kappa 2}(i)\bigr)\right)^2 \bigg/ (N-3) \qquad \text{(B1)}$$

, where $f_e$ represents the phase space densities of electrons from the measurements of FPI and FEEPS, $f_e^{\kappa 1}$ and $f_e^{\kappa 2}$ stand for the phase space densities from the two Kappa distribution functions and $N$ denotes the number of data points, which is 34 for our study.

## Acknowledgments.

This work is supported by NASA Grants 80NSSC23K1020, 80NSSC24K0069 and 80NSSC24K0452. AWS was supported by NERC Independent Research Fellowship NE/W009129/1. This research was supported by the NASA MMS in association with NASA contract NNG04EB99C at Southwest Research Institute (SwRI). Institut de Recherche en Astrophysique et Planétologie (IRAP) contributions to MMS FPI were supported by Centre National d'Études Spatiales (CNES) and Centre National de la Recherche Scientifique (CNRS). We thank the MMS team for data access and support. Weijie Sun thanks Dr. Shichen Bai from Shandong University, Weihai, for the helpful discussions.

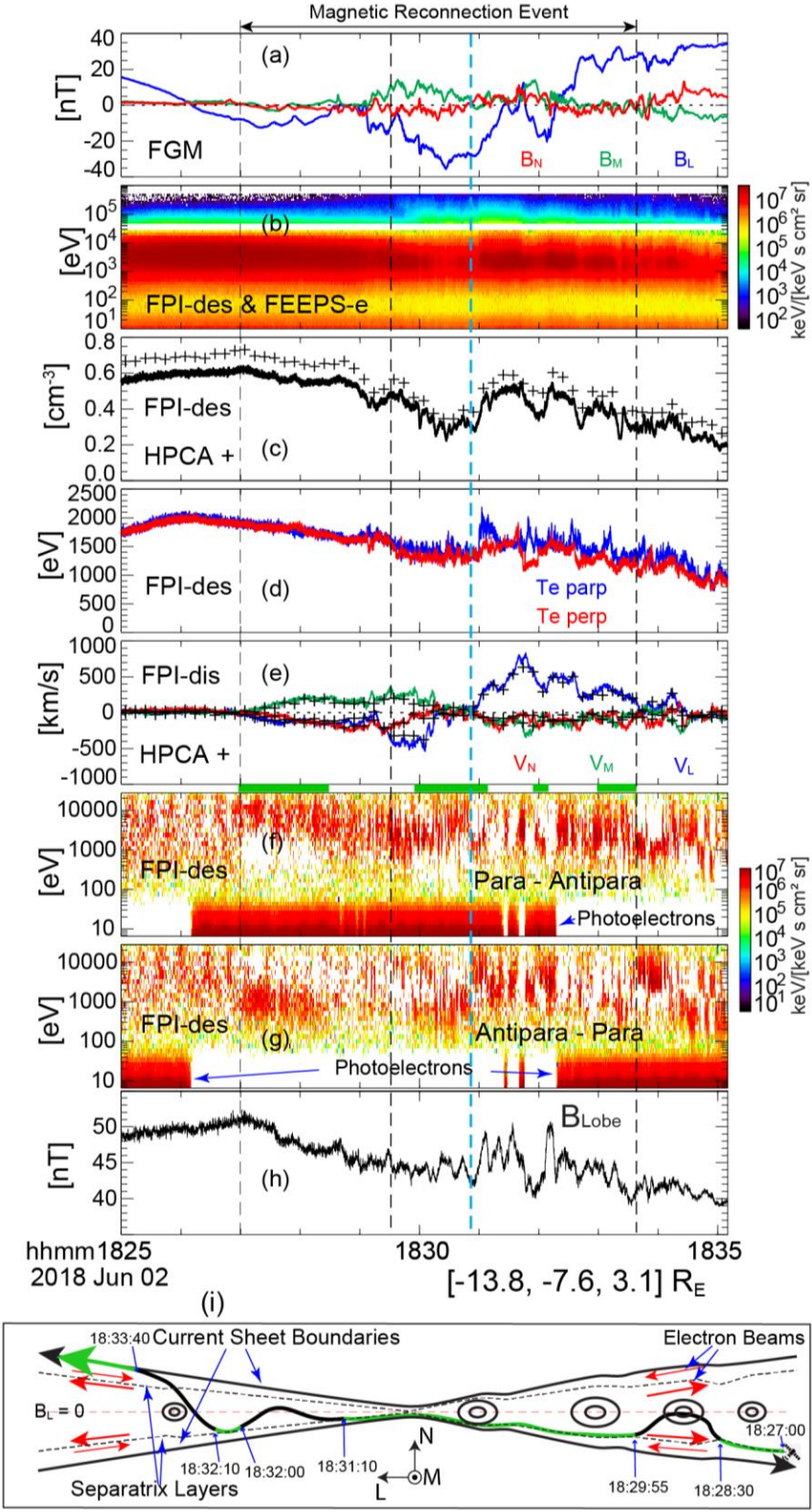

**Figure 1.** Overview of a magnetic reconnection event observed by MMS1 on 2 June, 2018 in Earth's magnetotail plasma sheet. (a) Magnetic field components in the local coordinate of the cross-tail neutral sheet, $B_L$ (blue), $B_M$ (green), $B_N$ (red). (b) Energy spectrogram for electrons

with energies from ~10 eV to 560 keV from FPI and FEEPS. (c) Ion density from HPCA and electron density from FPI. (d) Electron temperature from FPI, perpendicular ($T_{eperp}$, red) and parallel ($T_{eparp}$, blue) components. (e) Ion bulk velocity measured by FPI and HPCA, solid lines are from FPI with $V_L$ (blue), $V_M$ (green), $V_N$ (red), and crosses are from HPCA. (f) and (g) are energy spectra of electrons from FPI with electrons moving parallel to magnetic field subtracting those moving anti-parallel and *vice versa*, respectively. (h) The magnetic field intensity in the lobe derived through pressure balance. (i) A schematic figure of MMS's trajectory of the magnetic reconnection event. The green segments represent the separatrix regions, corresponding to time intervals represented by green bars between panels (e) and (f). The first and fourth vertical dashed lines indicate the start and the end of the reconnection event. The second vertical dashed line indicates the sharp enhancement in ion bulk velocity, while the third dashed line marks the primary X-line location.

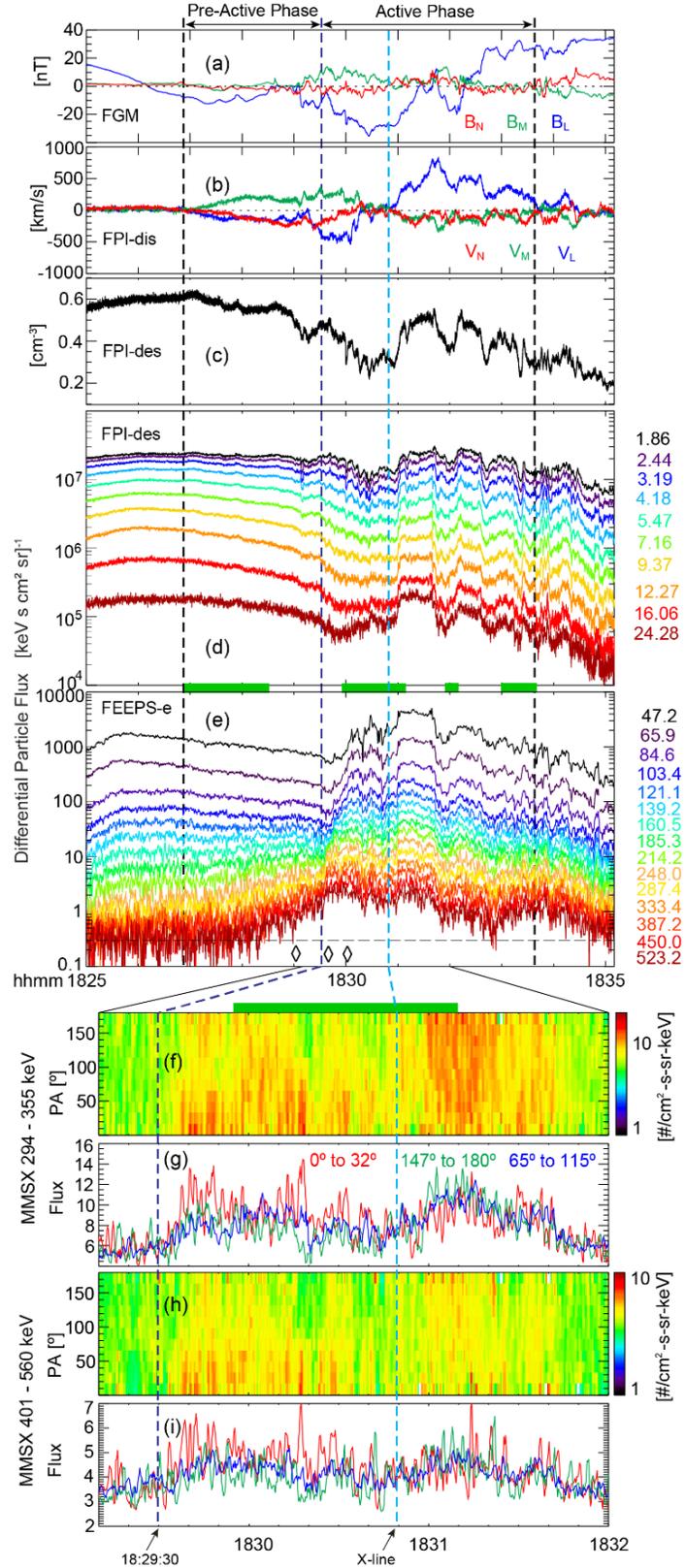

**Figure 2.** Differential particle flux variations of electrons with energies from ~ 1.86 to 523.2 keV from MMS1 for the reconnection event. (a) and (b) Magnetic field and ion bulk velocity

under the local coordinate of the neutral sheet. (c) Electron density. (d) and (e) Electron flux intensities measured by FPI and FEEPS. The horizontal line in (e) represents the background flux of the 523.2 keV energy channel. (f and g) Pitch angle distribution and fluxes in parallel (0° to 32°, red), perpendicular (65° to 115°, blue) and anti-parallel (147° to 180°, green) of electrons with energies between 294 to 355 keV. (h) and (i) are for electrons with energies between 401 to 560 keV, in a similar format to (f) and (g).

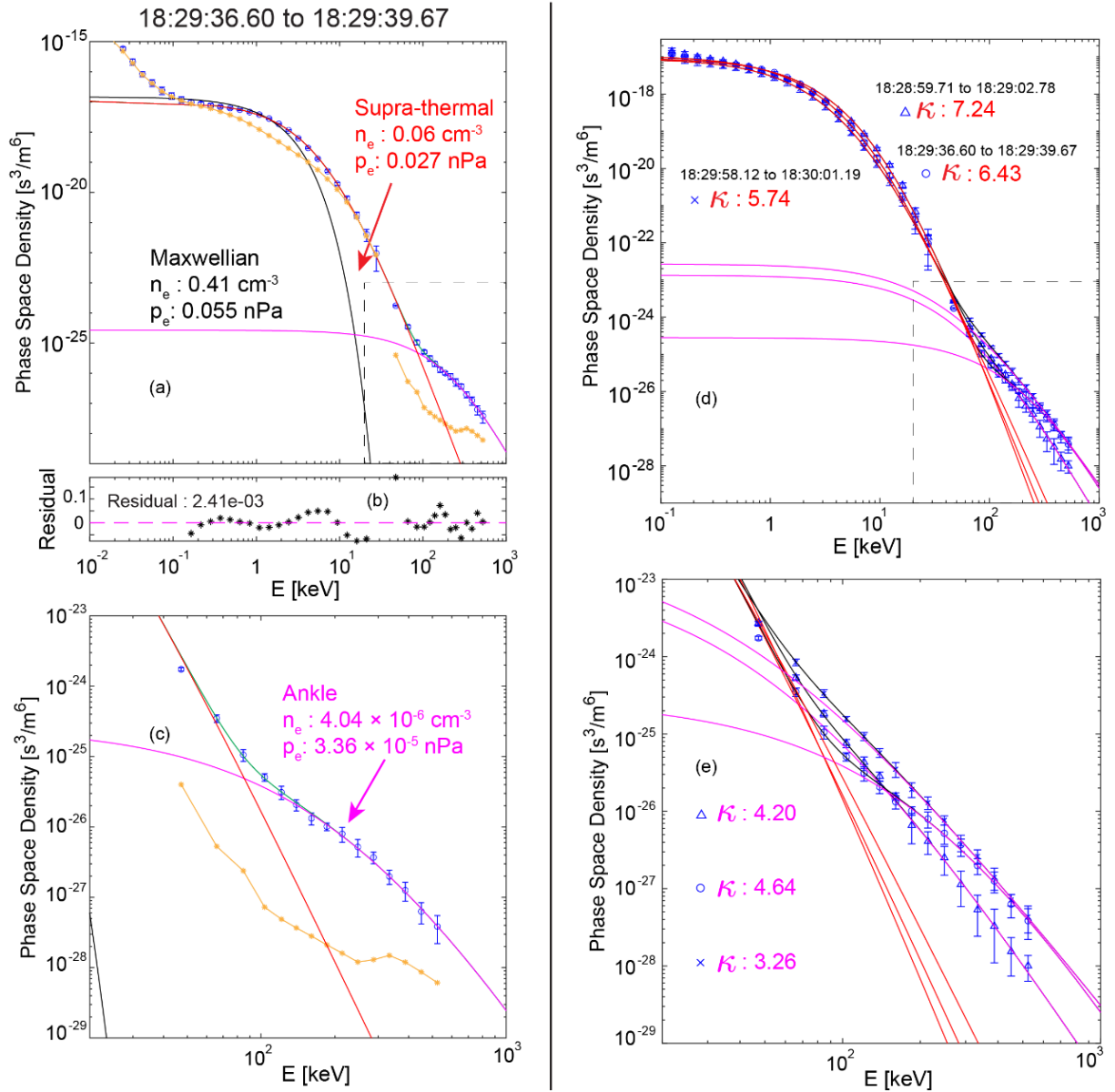

**Figure 3.** (a to c) Kappa function modelling of electron phase space densities obtained from 18:29:36.6 to 18:29:39.67 UTC. The blue dots with error bars represent measured phase space densities. The red and magenta lines correspond to two Kappa modellings for electrons, respectively. The black line represents the derived Maxwellian function for the relatively low energy Kappa function. Residuals in (b) are normalized (see Appendix B for details). The orange lines indicate the one-count-level phase space densities for FPI and FEEPS. (d) and (e) Three spectral distributions are shown. One is during pre-active phase (18:28:59.71 to 18:29:02.78 UTC, marked as triangles), one is during the energy dispersion (18:29:36.6 to 18:29:39.67 UTC, marked in circles), and one is in the separatrix layer during active phase (18:29:58.12 to 18:30:01.19 UTC, marked in crosses). Their locations are marked as the diamonds in Figure 2e. Relativistic effects were considered during conversion to phase space densities (Hilmer, Ginet, & Cayton 2000).

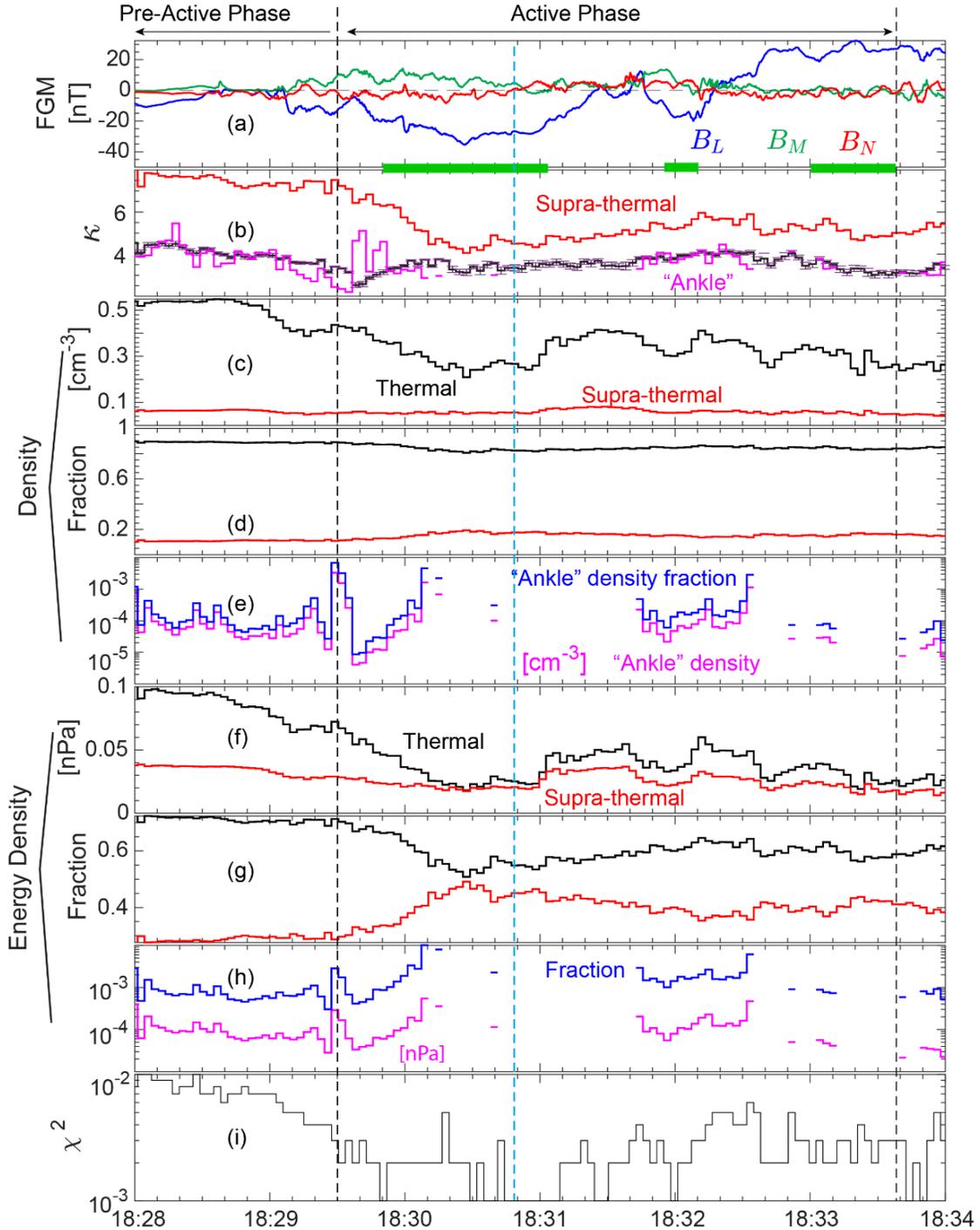

**Figure 4.** Number and energy densities of thermal, supra-thermal and relativistic "Ankle" electrons between 18:28:00 and 18:34:00 UTC. (a) Magnetic field components ($B_L$ in blue, $B_M$ in green, $B_N$ in red). (b) Spectral slope indices ($\kappa$) of supra-thermal (red) and the "ankle" (magenta and black) electrons. The red and magenta indices are obtained through the Kappa distributions and the black indices are obtained from power law distributions. (c) Densities of thermal (black) and supra-thermal (red) electrons. (d) Density fractions of thermal and supra-thermal electrons. (e) Density (blue) and density fraction (magenta) of the "Ankle" electrons. (f to h) are similar to (c to e), but are energy densities for thermal, supra-thermal and relativistic electrons. (i) Normalized residuals from the function modellings.

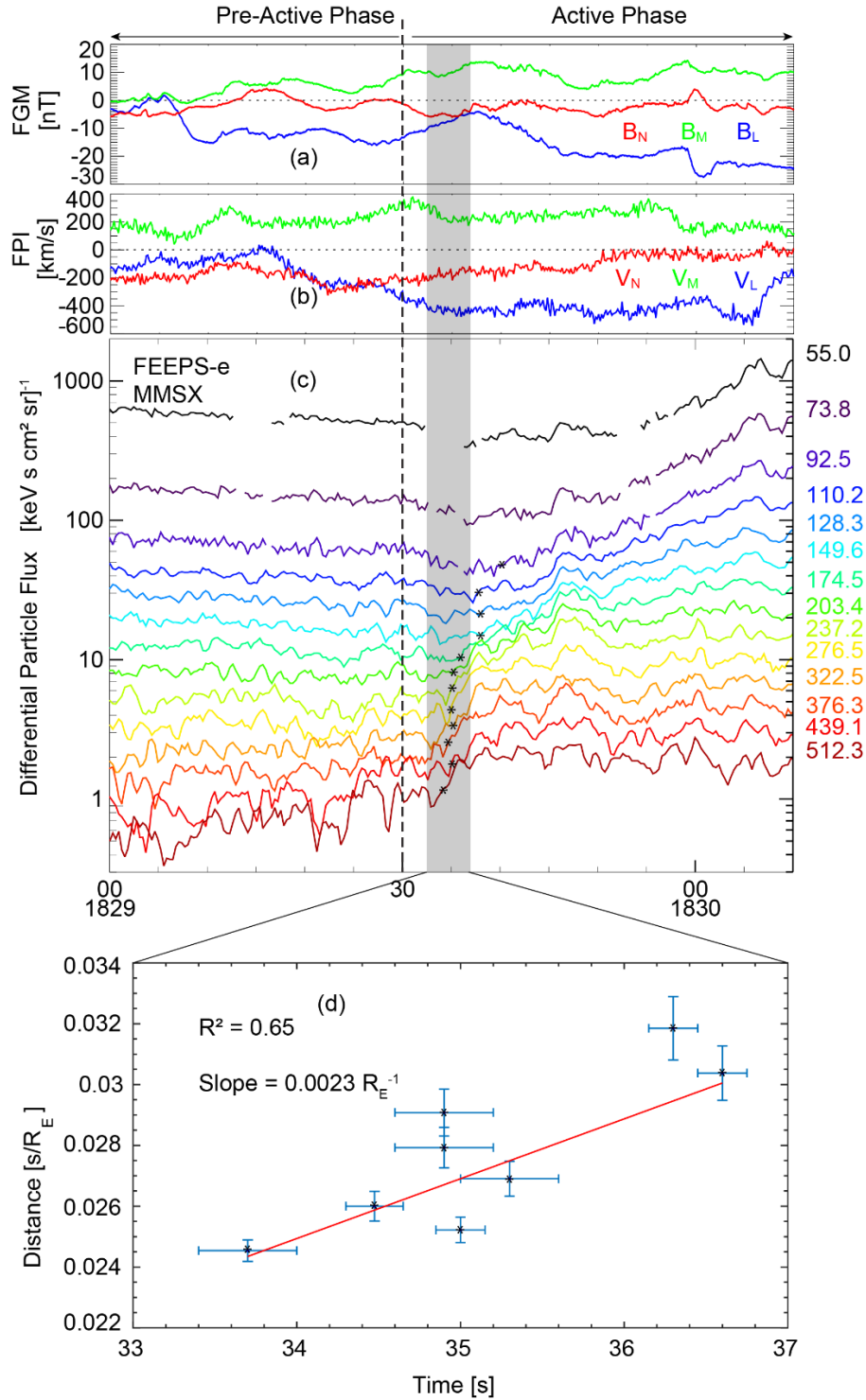

**Figure 5.** Overview of the enhancements of electrons with energies from ~ 47.2 to 523.2 keV at ~ 18:29:30 UTC when ion bulk velocity suddenly increased. (a) and (b) Magnetic field and ion bulk velocity in the local coordinate. (c) Electron flux intensities measured by FEEPS. The asterisks indicate the time when electron fluxes start to sharply increase in each energy channel.

(d) The fitting between $t$ and $1/V_{para}(t)$. The fitting includes the asterisks during the interval that MMS moved towards the neutral sheet as indicated by the shaded region. The $V_{para}$ is the parallel velocity of the dispersed electrons.

# References

Angelopoulos, V., Artemyev, A., Phan, T. D., & Miyashita, Y. 2020, Nature Physics

Angelopoulos, V., et al. 2008, Science, 321, 931

Bai, S.-C., et al. 2019, Journal of Geophysical Research: Space Physics, 124, 7494

Baker, D. N., Pulkkinen, T. I., Angelopoulos, V., Baumjohann, W., & McPherron, R. L. 1996, Journal of Geophysical Research: Space Physics, 101, 12975

Beyene, F., & Angelopoulos, V. 2024, Journal of Geophysical Research: Space Physics, 129, e2024JA032434

Blake, J. B., et al. 2016, Space Science Reviews, 199, 309

Büchner, J., & Zelenyi, L. M. 1989, Journal of Geophysical Research: Space Physics, 94, 11821

Burch, J. L., Moore, T. E., Torbert, R. B., & Giles, B. L. 2016a, Space Science Reviews, 199, 5

Burch, J. L., et al. 2016b, Science, 352, aaf2939

Chen, B., et al. 2020, Nature Astronomy, 4, 1140

Christon, S. P., Williams, D. J., Mitchell, D. G., Huang, C. Y., & Frank, L. A. 1991, Journal of Geophysical Research: Space Physics, 96, 1

Cohen, I. J., Turner, D. L., Mauk, B. H., Bingham, S. T., Blake, J. B., Fennell, J. F., & Burch, J. L. 2021, Geophysical Research Letters, 48, e2020GL090087

Contopoulos, I. 2005, Astronomy and Astrophysics, 442, 579

Dahlin, J. T., Drake, J. F., & Swisdak, M. 2014, Physics of Plasmas, 21, 092304

Demoulin, P., van Driel-Gesztelyi, L., Schmieder, B., Hemoux, J. C., Csepura, G., & Hagyard, M. J. 1993, Astronomy and Astrophysics, 271, 292

Denton, R. E., et al. 2018, Journal of Geophysical Research: Space Physics, 123, 2274

Drake, J. F., Swisdak, M., Che, H., & Shay, M. A. 2006, Nature, 443, 553

Duncan, R. C., & Thompson, C. 1992, The Astrophysical Journal, 392, L9

Egedal, J., Daughton, W., & Le, A. 2012, Nature Physics, 8, 321

Ellison, D. C., & Ramaty, R. 1985, The Astrophysical Journal, 298, 400

Genestreti, K. J., et al. 2023, Journal of Geophysical Research: Space Physics, 128, e2023JA031760

Genestreti, K. J., et al. 2018, Journal of Geophysical Research: Space Physics, 0

Gershman, D. J., et al. 2017, Journal of Geophysical Research: Space Physics, 122, 11

Gershman, D. J., et al. 2024, Space Science Reviews, 220, 7

Gonzalez, W., & Parker, E. 2016, Magnetic Reconnection: Concepts and Applications (Cham Heidelberg New York Dordrecht London: Springer)

Gonzalez, W. D., Tsurutani, B. T., Gonzalez, A. L. C., Smith, E. J., Tang, F., & Akasofu, S.-I. 1989, Journal of Geophysical Research: Space Physics, 94, 8835

Gray, P. C., & Lee, L. C. 1982, Journal of Geophysical Research: Space Physics, 87, 7445

Hakobyan, H., Philippov, A., & Spitkovsky, A. 2023, The Astrophysical Journal, 943, 105

Hilmer, R. V., Ginet, G. P., & Cayton, T. E. 2000, Journal of Geophysical Research: Space Physics, 105, 23311

Hoshino, M., Mukai, T., Terasawa, T., & Shinohara, I. 2001, Journal of Geophysical Research: Space Physics, 106, 25979

Imada, S., Hoshino, M., & Mukai, T. 2008, Journal of Geophysical Research: Space Physics, 113

Ji, H., Daughton, W., Jara-Almonte, J., Le, A., Stanier, A., & Yoo, J. 2022, Nature Reviews Physics, 4, 263

Jiang, K., et al. 2021, Geophysical Research Letters, 48, e2021GL093458

Li, X., Guo, F., Li, H., & Li, G. 2017, The Astrophysical Journal, 843, 21

Li, X., Guo, F., Li, H., Stanier, A., & Kilian, P. 2019, The Astrophysical Journal, 884, 118

Lyubarsky, Y., & Kirk, J. G. 2001, The Astrophysical Journal, 547, 437

Lyutikov, M. 2003, Monthly Notices of the Royal Astronomical Society, 346, 540

Ma, W., Zhou, M., Zhong, Z., & Deng, X. 2024, Journal of Geophysical Research: Space Physics, 129, e2023JA032270

Moré, J. J. 1977. in Numerical analysis: proceedings of the biennial Conference held at Dundee, June 28–July 1, 1977, The Levenberg-Marquardt algorithm: implementation and theory (Springer), 105

Nagai, T., Shinohara, I., Fujimoto, M., Hoshino, M., Saito, Y., Machida, S., & Mukai, T. 2001, Journal of Geophysical Research: Space Physics, 106, 25929

Øieroset, M., Lin, R. P., Phan, T. D., Larson, D. E., & Bale, S. D. 2002, Physical Review Letters, 89, 195001

Olbert, S. 1968. in Physics of the Magnetosphere, Summary of Experimental Results from M.I.T. Detector on IMP-1, eds. R. L. Carovillano, J. F. McClay, & H. R. Radoski (Dordrecht: Springer Netherlands), 641

Oka, M., Krucker, S., Hudson, H. S., & Saint-Hilaire, P. 2015, The Astrophysical Journal, 799, 129

Oka, M., et al. 2018, Space Science Reviews, 214, 82

Oka, M., et al. 2022, Physics of Plasmas, 29

Phan, T. D., et al. 2018, Nature, 557, 202

Pollock, C., et al. 2016, Space Science Reviews, 199, 331

Qi, Y., et al. 2024, The Astrophysical Journal Letters, 962, L39

Russell, C. T., et al. 2016, Space Science Reviews, 199, 189

Shi, Q. Q., et al. 2006, Geophysical Research Letters, 33

Shi, Q. Q., et al. 2005, Geophysical Research Letters, 32

Shi, Q. Q., et al. 2019, Space Science Reviews, 215, 35

Shibata, K. 1998, Astrophysics and Space Science, 264, 129

Shue, J.-H., et al. 1998, Journal of Geophysical Research: Space Physics, 103, 17691

Slavin, J. A., Smith, M. F., Mazur, E. L., Baker, D. N., Iyemori, T., Singer, H. J., & Greenstadt, E. W. 1992, Geophysical Research Letters, 19, 825

Sonnerup, B. U. Ö. 1979, in Space Plasma Physics: The study of Solar System plasmas (Washington, D. C.: National Academy of Sciences), 879

Sonnerup, B. U. Ö., & Cahill Jr., L. J. 1967, Journal of Geophysical Research (1896-1977), 72, 171

Sonnerup, B. U. Ö., & Scheible, M. 1998, in Analysis methods for multi-spacecraft data, eds. G. Paschmann, & P. W. Daly (Noordwijk, Netherlands.: ESA Publication), 185

Sun, W., et al. 2022, Journal of Geophysical Research: Space Physics, 127, e2022JA030721

Tanabashi, M., et al. 2018, Physical Review D, 98, 030001

Torbert, R. B., et al. 2018, Science, 362, 1391

Vasyliunas, V. M. 1968, Journal of Geophysical Research (1896-1977), 73, 2839

Wang, C.-P., Lyons, L. R., Weygand, J. M., Nagai, T., & McEntire, R. W. 2006, Journal of Geophysical Research: Space Physics, 111

Yamada, M., Kulsrud, R., & Ji, H. 2010, Reviews of Modern Physics, 82, 603

Young, D. T., et al. 2016, Space Science Reviews, 199, 407

Zhong, Z. H., et al. 2020, Geophysical Research Letters, 47, e2019GL085141